\documentclass[prc,twocolumn,epsfig,nofootinbib,floatfix]{revtex4}
\usepackage{graphics}
\usepackage{epsfig}
\usepackage{amsfonts}
\usepackage{amsmath}
\usepackage{bm}

\usepackage{color}

\begin{document}
\title{Relation of E1 pygmy and toroidal resonances}
\author{
V.O. Nesterenko$^{1}$, A. Repko$^{2}$, P.-G. Reinhard $^{3}$, and J. Kvasil$^{2}$}
\affiliation{$^{1}$ \it
Laboratory of Theoretical Physics, Joint Institute for Nuclear
Research, Dubna, Moscow region, 141980, Russia}
\email{nester@theor.jinr.ru}
\affiliation{$^{2}$ \it Institute of Particle and Nuclear Physics, Charles
University, CZ-18000, Prague, Czech Republic}
\affiliation{$^{3}$ \it Institut f\"ur Theoretische Physik II,
Universit\"at Erlangen, D-91058, Erlangen, Germany}

\begin{abstract}
A possible relation of the low-lying E1 (pygmy
resonance) and toroidal strengths is analyzed by using
  Skyrme-RPA results for the strength functions, transition densities
  and current fields in $^{208}$Pb.  It is shown that the irrotational
  pygmy motion can appear as a local manifestation of the collective
  vortical toroidal dipole resonance (TDR) at the nuclear surface. The
  RPA results are compared to unperturbed (1ph) ones.
\end{abstract}
\maketitle
\section{Introduction}
\label{intro}

Last years there is a high interest in low-lying E1 strength,
  often called pygmy dipole resonance (PDR)
\cite{Paar07,Savran13}. The PDR strength is related to nuclear
matter properties (symmetry energy, incompressibility, effective
masses) \cite{Rei13f} and so can deliver useful complementing
information for building the isospin-dependent part of the nuclear
equation of state and for astrophysical applications.

The PDR is fragmented into several peaks \cite{Rei13f} which can
be separated into isoscalar (T=0) low-energy and isospin-mixed
(T=0,1) high-energy parts \cite{En10}. In our recent study
\cite{Repko13}, we suggested that the low-energy isoscalar part of
the PDR can be interpreted as a peripheral fraction of the
isoscalar toroidal dipole resonance (TDR) \cite{Du75,Se81}.  The
statement was based on a careful analysis of the E1(T=0) strength
functions, transition densities and current fields in $^{208}$Pb,
performed within the Skyrme random-phase-approximation (RPA).

This suggestion is not so obvious if one views the PDR as a
collective mode built from oscillations of the neutron excess
against the nuclear core N=Z \cite{Paar07}, i.e. as a fully {\it
irrotational} mode occurring in nuclei with N>Z.  Instead, the TDR
is a {\it vortical} motion \cite{Repko13,Du75}.  Moreover, the TDR
seems to be the main carrier of the nuclear dipole vorticity
\cite{Rein14}. This resonance should exist in all nuclei
independent of its neutron content. After extraction of the
center-of-mass corrections (cmc), the TDR dominates in low-energy
E1(T=0) excitations and, probably, constitutes the low-energy part
of the isoscalar giant dipole resonance (ISGDR) observed in
$(\alpha,\alpha')$ reaction \cite{Uchida04}. The TDR transition
operator is \cite{Kv11}
\begin{equation}
\label{TM_curl} \hat M_{\text{TMR}}(E1\mu)
\propto \int d^3r [r^3-\frac{5}{3}r \langle r^2\rangle_0] \vec
Y_{11\mu}(\hat{\vec r}) \cdot [\vec{\nabla}\!\times\!\hat{\vec
j}(\vec r]) \; ,
\end{equation}
where $\hat{\vec j}(\vec r)$ is the operator of the  nuclear
current, $\vec{Y}_{11\mu}(\hat{\vec r})$ is the vector spherical
harmonic. The cmc includes $\langle r^2\rangle_0 = \int\:d^3r
\:r^2 \rho_0(\vec{r})/A$ with $\rho_0(\vec{r})$ being the g.s.
density. The operator is determined by the curl of the nuclear
current and so is vortical.

In this paper, we present additional arguments that, despite the
PDR/TDR differences (e.g. irrotational vs vortical flow), the
isoscalar PDR may rather be considered as a local manifestation of
the general toroidal motion \cite{Rei13f}. The collectivity of the
TDR is inspected in detail. Like in the study \cite{Repko13}, the
analysis is done for $^{208}$Pb in the framework of the 1d RPA
with the Skyrme force SLy6 \cite{SLy6}.

\section{Discussion}
\label{sec-1}

 In Figure 1, the dipole strength functions are shown for the relevant
 resonances. The energy-dependent Lorentz smoothing \cite{Kva_IJMP_11}
 is used to roughly simulate the dissipation effects (coupling to
 complex configuration and continuum) beyond our RPA description.
 Panel (a) indicates the regions of PDR and giant dipole resonance (GDR).
 The PDR has two peaks at 7.5 and 10.3 MeV, which is in agreement with
 the previous calculations using relativistic mean field (RMF)
 \cite{Vr01} or Skyrme functional \cite{Rei13f}. In the panels
 (b, c), the isoscalar TDR and its high-energy counterpart, the
 compression dipole resonance (CDR) \cite{Ha77,St82}, are depicted. It
 is seen that the TDR lies at 6-9 MeV, i.e. just at the energy
 interval where the isoscalar PDR is located. Moreover the TDR energy
 almost coincides with the position of the PDR peak at 7.5 MeV. So the
 PDR and TDR are related. Also we see here a small fraction of CDR.

Panel (b) demonstrates a considerable difference between RPA and
unperturbed (1ph) TDR responses. The residual isoscalar
interaction noticeably downshifts the strength and leads to the
strong peak at 7.5 MeV. Actually this peak embraces two RPA states
at 7.4 and 7.7 MeV. The states emerge from superposition of
several 1ph components with maximal ones exhausting 67$\%$ and
23$\%$ of the norm, respectively. So the TDR seems to be
collective.
\begin{figure}
\centering \resizebox{0.45\textwidth}{6.4cm} {\includegraphics
{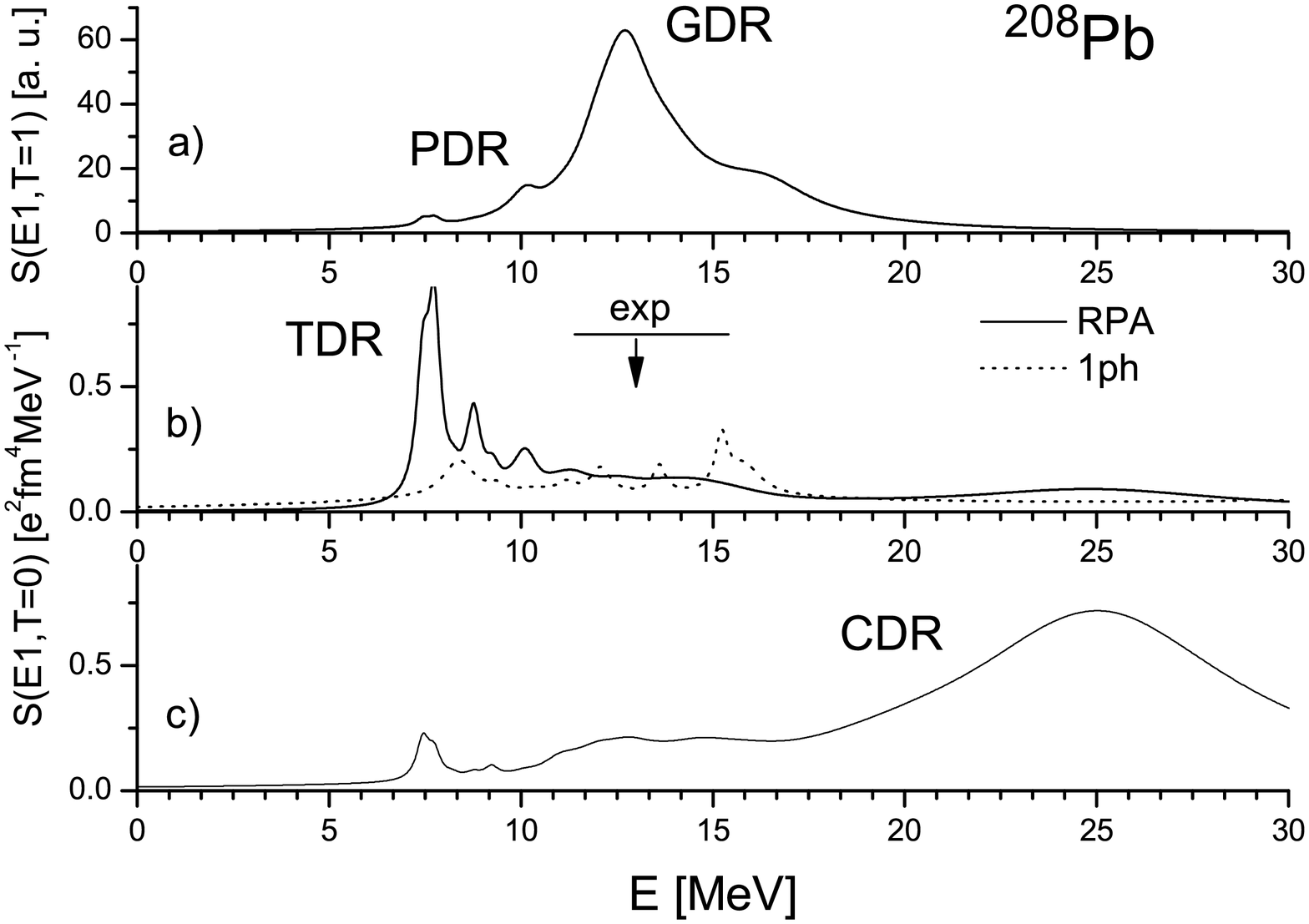}} \caption{Calculated strength functions: (a)
energy-weighted E1(T=1) GDR and PDR, (b) E1(T=0) TDR (RPA and
unperturbed 1ph), (c) E1(T=0) CDR. In (b), the experimental width
and energy of the ISGDR low-energy branch \protect\cite{Uchida04}
are denoted.} \label{fig-1}
\end{figure}

Panel (b) also shows that the calculated TDR lies noticeably below
the low-energy ISGDR branch observed in ($\alpha, \alpha'$)
reaction at 12-14 MeV \cite{Uchida04}. The same discrepancy takes
place in most of the theoretical studies \cite{Paar07}. Following
panel (c), perhaps not the TDR but the low-energy CDR bump was
observed.  Note also that the experiment \cite{Uchida04} has
inspected the excitation energy interval 8-35 MeV and so might
lose the strong and narrow TM peak at 7-8 MeV. Therefore,
additional experiments to search the isoscalar TDR are desirable.
\begin{figure}
\centering \resizebox{0.44\textwidth}{3.6cm}{
  \includegraphics{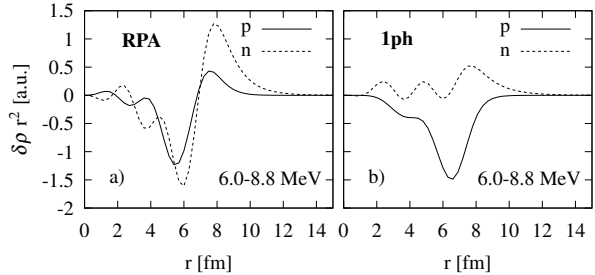}
} \caption{The calculated RPA (left) and unperturbed 1ph (right)
$r^2$-weighted proton and neutron TD summed at the energy interval
6.0-8.8 MeV \protect\cite{Repko13}.}
\label{fig-2}
\end{figure}

Strength functions, being integral observables, are too rough for
a thorough analysis of the subtle relation between TDR and PDR.
They should be complemented by more detailed observables:
transition densities (TD) and current transition densities (CTD)
visualizing the coordinate dependence of the nuclear flow. In
order to weaken the sensitivity of the results to individual
dipole states and so to focus on the underlying gross features of
flow, the TD/CTD strengths are summed over all RPA states in a
properly chosen energy interval. This is done here for the
interval 6.0-8.8 MeV embracing both T=0 PDR and TDR. To circumvent
the ambiguity in signs the RPA states, the TD/CTD  are calculated
with the corresponding isoscalar weights, see \cite{Repko13} for
details. In CTD the convection nuclear current is used.
\begin{figure}
\centering
\resizebox{0.42\textwidth}{6.8cm}{\includegraphics{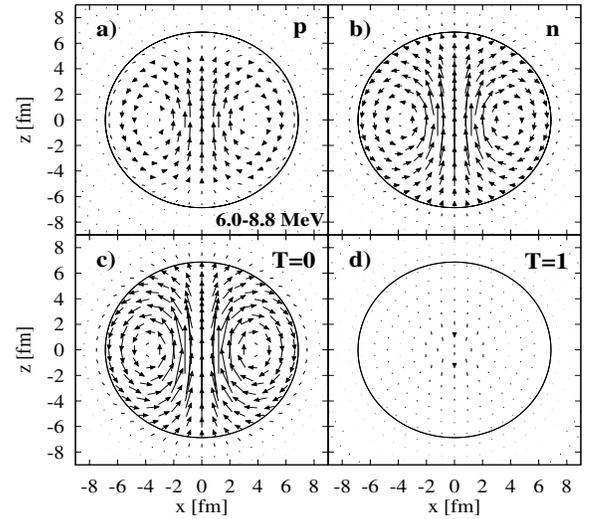}}
\caption{RPA proton (a), neutron (b), T=0 (c) and T=1 (d) CTD
summed at the energy interval 6.0-8.8 MeV.} \label{fig-3}
\end{figure}

The treatment of the PDR as oscillations of a neutron excess
against the nuclear N=Z core is often justified by the specific
picture of the PDR transition densities where neutron and protons
move by a similar manner in the nuclear interior while the neutron
motion strongly dominates at the nuclear surface \cite{Paar07}. In
Fig. 2(a), these are regions 4-7 fm and 7-11 fm, respectively. It
is also seen (plot (b)) that the pygmy-like picture is lost in the
summed 1ph TD.
\begin{figure}
\centering
\vspace{0.3cm}
\resizebox{0.42\textwidth}{6.8cm}{\includegraphics{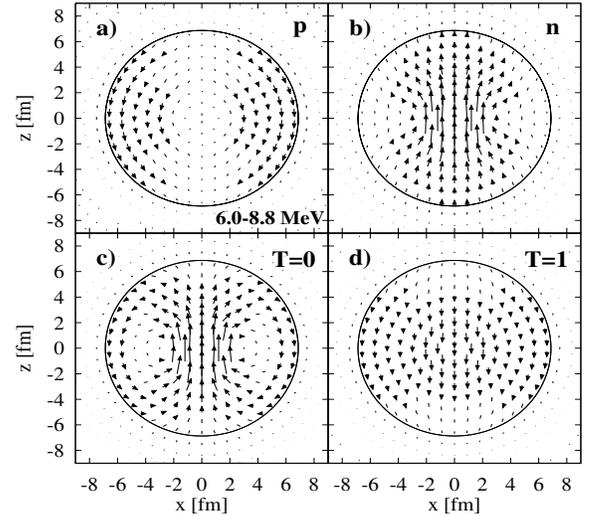}}
\caption{Unperturbed (1ph) proton (a), neutron (b), T=0 (c) and
T=1 (d) CTD summed at the energy interval 6.0-8.8 MeV.}
\label{fig-4}
\end{figure}

However TD illustrates only radial motion while a thorough flow
image needs an angular view as well. The latter is provided by the
CTD given in Figs. 3 and 4 for the RPA and 1ph cases,
respectively. Fig. 3 shows that neutron and, in a less extent,
proton motions exhibit a typical toroidal flow (to be compared to
the schematic TDR view in Fig. 4(a)). Both motions are in phase
and give altogether a strong isoscalar TDR while the isovector
field is suppressed. So, following the current fields, the E1 RPA
excitations at 6.0-8.8 MeV (location of T=0 PDR) are mainly of
isoscalar vortical toroidal nature. A similar result for CTD was
obtained earlier within the quasiparticle-phonon model
\cite{Ry02}.
\begin{figure}\centering
\resizebox{0.47\textwidth}{3cm}
{\includegraphics{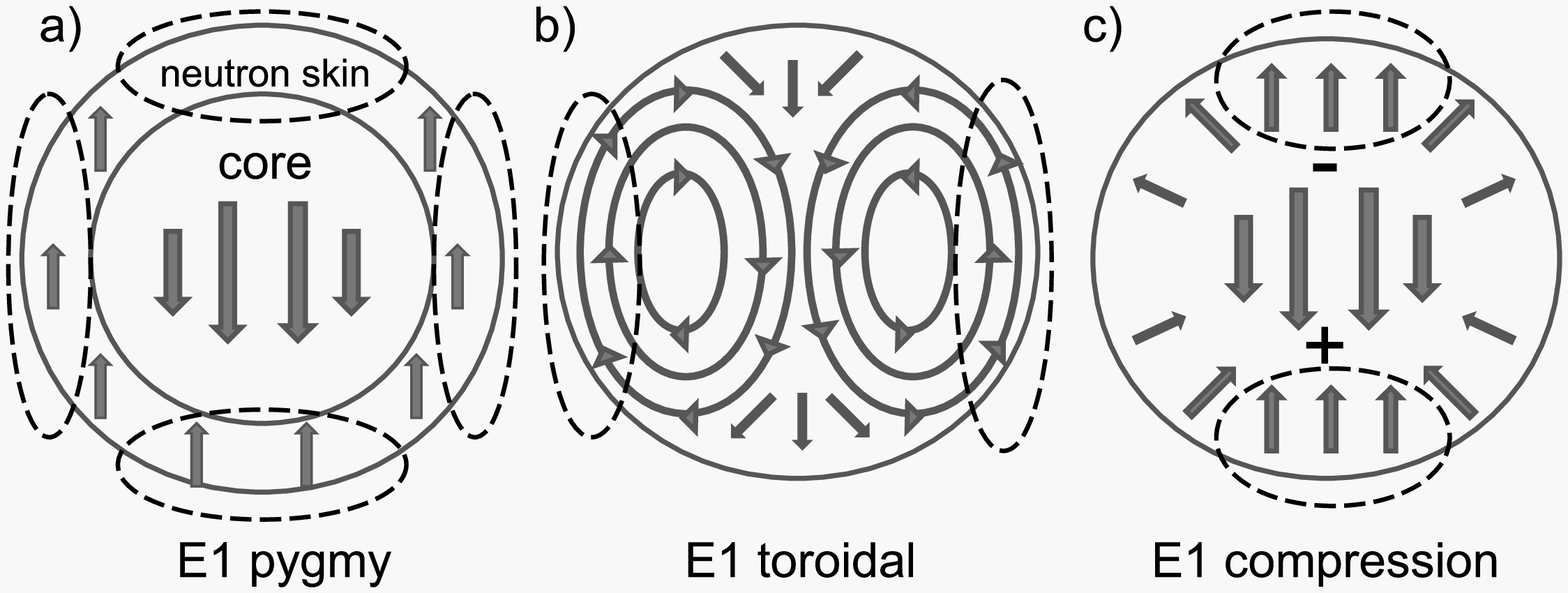}} \caption{Schematic
velocity fields for the PDR (a), TDR (b), and CDR (c)
\protect\cite{Repko13}. The driving field is directed along
z-axis. In the plot (c), the compression (+) and decompression (-)
regions are marked. The dashed ovals show similar local peripheral
flows in the PDR and TDR/CDR.} \label{fig-5}
\end{figure}

We also see clear signs of toroidal flow in unperturbed neutron
and proton current fields depicted in Fig. 4. However, unlike the
RPA case, the T=1 GDR-like field is comparable here in magnitude
with the T=0 toroidal field. This difference between RPA and 1ph
cases, together with those in Figs. 1 and 2, allows to assign some
collectivity to the T=0 TDR.

The natural question is how to conform rather contradicting
irrotational PDR and vortical TDR pictures? For this aim, it is
instructive to consider schematic views \cite{Repko13} of the PDR,
TDR, and CDR, given in Fig. 5. These idealized pictures indicate
that PDR and CDR are indeed irrotational while the TDR is fully
vortical. This is additionally confirmed by expressions for their
velocities \cite{Kv11}: $\vec{v}_{\rm{PDR,GDR}}=\vec{\nabla} (r
Y_{1\mu})$, $\vec{v}_{\rm{CDR}}=\vec{\nabla} (r^3 Y_{1\mu})$, and
$\vec{v}_{\rm{TDR}}= r^2 \vec{Y}_{12\mu}$. However, the peripheral
motion of the neutron excess in the PDR is similar to the
peripheral nucleon motions in the TDR (left/right sides) and CDR
(up/bottom sides), as indicated in Fig. 5 by dash ovals. In other
words, despite the complete toroidal motion is vortical, its
peripheral left and right parts look irrotational. Hence the
principle difference between the irrotational PDR and vortical TDR
becomes locally insignificant and the PDR can indeed be seen as a
local manifestation of the TDR.

The next PDR/TDR difference is that the PDR exists only in nuclei
with the neutron excess while the TDR is pertinent to all nuclei.
However this also does not contradict our treatment since in
nuclei with N>Z, the peripheral flow in TDR and CDR should be
dominated by neutrons. Anyway our {analysis should still be
further checked by systematic calculations for PDR/TDR in
different mass regions. It should be verified if indeed the PDR
excitations always share the energy region with the TDR and
exhibit the toroidal current fields. Though the coupling with
complex configuration is essential in the PDR region \cite{En10},
it should not affect much the summed current fields (as follows
from the similarity of our fields to those from \cite{Ry02}).

Though E1 excitations at 6.0-8.8 MeV seem to be mainly  toroidal,
they can also have minor contributions from other modes (CDR, GDR
tail, individual 1ph configurations) which are not well visible in
the current fields but follow from the analysis of the strength
functions. So, most probably, these excitations are a complicated
mixture of various coupled modes with their own doorway states.
Depending on the experimental reaction, only the modes with proper
doorway states are directly excited while others are involved only
due to the coupling. Then, depending on the experimental probe,
the dipole states can respond as the PDR in photoabsorption or T=0
TDR/CDR in $(\alpha,\alpha')$.

In summary, following our analysis, the isoscalar part of the PDR
can be interpreted (fully or partly) as a  manifestation of the
toroidal resonance at the nuclear surface. The TDR demonstrates
some collectivity.

 The work was partly supported by the DFG grant RE 322/14-1,
Heisenberg-Landau (Germany-BLTP JINR), and Votruba-Blokhincev
(Czech Republic-BLTP JINR) grants. P.-G.R. is grateful for the
BMBF support under Contracts No, 06 DD 9052D and 06 ER 9063. The
support of the Czech Science Foundation (P203-13-07117S) is
appreciated.

\end{document}